\documentclass[10pt]{article}
\usepackage{graphics}
\usepackage{graphicx}
\usepackage{xcolor}
\usepackage{hyperref} 

\begin{document}
\noindent {\large\bf High precision wavefront control in point spread function engineering for single emitter localization} \\[2ex] 

\noindent M. Siemons,$^{1,2}$ C. N. Hulleman,$^{1}$ R. {\O}. Thorsen,$^{1}$ C. S. Smith,$^{3}$ and S. Stallinga$^{1,*}$\\[2ex]

\noindent $^{1}$ Department of Imaging Physics, Delft University of Technology, Delft, The Netherlands\\
\noindent $^{2}$ Currently with Cell Biology, Department of Biology, Faculty of Science, Utrecht University, Utrecht, The Netherlands\\
\noindent $^{3}$ Centre for Neural Circuits and Behaviour, University of Oxford, Oxford, United Kingdom\\
\noindent $^{*}$ s.stallinga@tudelft.nl 



\begin{abstract}
Point spread function (PSF) engineering is used in single emitter localization to measure the emitter position in 3D and possibly other parameters such as the emission color or dipole orientation as well. Advanced PSF models such as spline fits to experimental PSFs or the vectorial PSF model can be used in the corresponding localization algorithms in order to model the intricate spot shape and deformations correctly. The complexity of the optical architecture and fit model makes PSF engineering approaches particularly sensitive to optical aberrations. Here, we present a calibration and alignment protocol for fluorescence microscopes equipped with a spatial light modulator (SLM) with the goal of establishing a wavefront error well below the diffraction limit for optimum application of complex engineered PSFs. We achieve high-precision wavefront control, to a level below 20 m$\lambda$ wavefront aberration over a 30 minute time window after the calibration procedure, using a separate light path for calibrating the pixel-to-pixel variations of the SLM, and alignment of the SLM with respect to the optical axis and Fourier plane within 3~$\mu$m ($x/y$) and 100 $\mu$m ($z$) error. Aberrations are retrieved from a fit of the vectorial PSF model to a bead $z$-stack and compensated with a residual wavefront error comparable to the error of the SLM calibration step. This well-calibrated and corrected setup makes it possible to create complex `3D+$\lambda$' PSFs that fit very well to the vectorial PSF model. Proof-of-principle bead experiments show precisions below 10~nm in $x$, $y$, and $\lambda$, and below 20~nm in $z$ over an axial range of 1~$\mu$m with 2000 signal photons and 12 background photons.
\end{abstract}


\section{Introduction}
The diffraction limit to resolution is overcome in Single Molecule Localization Microscopy (SMLM) by estimating the location of individual molecules from sparsely distributed emission spots across the field of view of the camera \cite{Betzig2006,Rust2006,Hess2006,Hell2007}. The achievable resolution is limited by the localization precision and the density of fluorescent labels and can be on the order of several tens of nanometers in practice \cite{Nieuwenhuizen2013}. Several groups have extended this technique to 3D localization. Estimation of the axial position of the molecules is made possible by adding a cylindrical lens to the optical path \cite{Kao1994,Holtzer2007,Huang2008} or by using bi-plane or multi-focus imaging \cite{Toprak2007,Ram2008,Juette2008}. The elongation of the spot (astigmatism) or the difference in spot size (bi-plane imaging) encodes for the axial position. The addition of a Spatial Light Modulator (SLM) incorporated with a 4F relay system to the imaging light path enables the design of a broader class of Point Spread Functions (PSFs), opening up the possibility to optimize the localization performance of the microscope \cite{Shechtman2014}. Notable proposals in the literature are single and double helix PSFs derived from Gauss-Laguerre modes \cite{Lew2011,Pavani2008,Pavani2009,Grover2010}, or made with annular zones with increasing helical charge \cite{Prasad2013,Roider2014}, saddle point or tetrapod designs \cite{Shechtman2015}, which use higher orders of astigmatism in addition to primary astigmatism, phase-ramps \cite{Baddeley2011} and self-bending beams \cite{Jia2014}. Interferometric imaging can also be used to establish high-performance axial localization, but requires a complex 4$\pi$ optical setup \cite{Shtengel2009}. 

Other properties of the fluorescent molecules could be of interest next to the 3D-position, such as the emitter dipole orientation \cite{Backlund2014} and the wavelength of the emitted light. The latter is relevant for imaging multiple protein species in a specimen that are labeled with fluorophores with different (excitation and) emission spectra. Encoding the emission color into the PSF shape enables multi-color imaging with a single imaging light path and a single camera, operating at the full field of view. This has been proposed by our group in \cite{Broeken2014} for 2D-localization and generalized in \cite{Smith2016} for 3D-localization. An alternative approach to `3D+$\lambda$' localization has been described in \cite{Shechtman2016} by Shechtman et al.. The key idea of \cite{Broeken2014,Smith2016} is to have the SLM function as a Diffractive Optical Element (DOE), which splits the emission spot into two or three sub-spots. These sub-spots correspond to the diffraction orders generated by the DOE, and the distance between the spots and their relative intensity provide information on the emission wavelength of the fluorophore. The shape of the repetitive zones of the DOE can be designed for making the shape of the sub-spots change with the axial position of the fluorophore, thereby enabling 3D-localization.

A common characteristic of all engineered PSFs is their complexity compared to the simple 2D focused spots, which must be represented in the PSF model that is used in the parameter fitting algorithm for estimating the 3D-position (and possibly the emission color or molecular orientation). Simplified PSF models such as the Gaussian model \cite{Stallinga2010}, the scalar diffraction based Airy model, the Gibson-Lanni model \cite{Gibson1992}, or effective models based on Hermite functions \cite{Stallinga2012} cannot meet this requirement. A solution is the use of an experimental reference PSF, or a spline fit of such a PSF as model PSF \cite{Kirshner2013,Tahmasbi2015,Babcock2017,Li2017}, or the use of one or multiple Look Up Tables (LUTs) to estimate the z-position \cite{Diezman2015}. We have shown previously that a vectorial PSF model can also be used for complex 3D and 3D+$\lambda$ engineered PSFs \cite{Smith2016}. It is known that the vectorial PSF model is the physically correct model for image formation in high-NA fluorescence imaging systems. Another common characteristic of complex engineered PSFs is the sensitivity to aberrations that perturb the designed PSF shape and in this way negatively affect precision and accuracy. In order to achieve precisions down to the Cram\'{e}r-Rao Lower Bound (CRLB), the best possible precision for an unbiased estimator, the aberration level of the optical system should be controlled to well within the diffraction limit (0.072$\lambda$ root mean square wavefront aberration) \cite{Smith2016}, a condition which is often not met in practice. Correction of aberrations using a deformable mirror or with the SLM that is present anyway for producing the engineered PSF is therefore required. The control parameters of the adaptive optics component can be set using image based metrics \cite{Debarre2007,Booth2015,Burke2015} or via measurement of the to-be corrected aberrations. The latter may be done via phase retrieval algorithms based on the introduction of phase diversity, often in the form of a through-focus bead scan. This has been implemented  in high numerical aperture microscope system \cite{Hanser2003}, in localization microscopy \cite{Liu2013} and used to improve the quality of a STED laser focus \cite{Kromann2012}. These algorithms rely on smoothing and apply constraints to suppress noise effects. The noise statistics however can be incorporated into a Maximum Likelihood Estimation (MLE) based phase retrieval algorithm, which makes it possible to assess the optimality of the retrieval process by means of a CRLB analysis \cite{Petrov2017}.

Here, we present a calibration and alignment protocol for fluorescence microscopes equipped with an SLM with the goal of establishing a wavefront error well below the diffraction limit, which is needed for an optimum application of complex engineered PSFs in single emitter localization. This high-precision wavefront control is combined with the use of the vectorial PSF model, both in the MLE-based phase retrieval algorithm for estimating the aberrations, and in the localization algorithm for estimating the emitter position in 3D, signal photon count, and background photon level (and possibly emission color). We report on proof-of-principle experiments on fluorescent beads for different 3D+$\lambda$ engineered PSFs for testing whether the proposed calibration protocol and fitting with a vectorial PSF model give rise to precisions matching the CRLB.

The outline of this paper is as follows. In section 2 the vectorial fitting routine and aberration retrieval is described, section 3 describes the experimental setup and the calibration and alignment techniques, section 4 shows the experimental results on the aberration retrieval and correction as well as on different 3D+$\lambda$ engineered PSFs, and section 5 summarizes the main conclusions of the paper.

\section{Theory}
\subsection{Vectorial PSF model for an emitter with non-zero size}
The vectorial PSF model \cite{Mortensen2010,Smith2016} is further refined by taking into account the non-zero size of the fluorescent beads we use in the experiments (175~nm and 200~nm), which induces a blurring on the scale of 1-2 pixels. We describe the PSF of such a bead as a PSF of a freely rotating single dipole emitter convoluted with a sphere $\bigcirc(\vec{r}) $:
\begin{equation}
H_{\mathrm{bead}}\left(\vec{r}\right) = \frac{N}{3}\sum_{l=x,y}\sum_{j=x,y,z} |w_{lj}\left(\vec{r}\right)|^2\otimes \bigcirc\left(\vec{r}\right) + \frac{b}{a^2} ,
\end{equation}
with $H_{\mathrm{bead}}\left(\vec{r}\right)$ the PSF, $N$ the total number of signal photons detected at the camera, $b$ the number of background photons per pixel and $a$ the pixel size of the camera pixel, $\bigcirc\left(\vec{r}\right) $ = 1 for $|\vec{r}|<r_\mathrm{bead}$ and 0 otherwise with $r_\mathrm{bead}$ the radius of the bead and $w_{lj}\left(\vec{r}\right)$ the electric field component in the image plane, which is given by:
\begin{equation}
w_{lj}\left(\vec{r}\right) = \frac{1}{c_n} \int_{|\vec{\rho}|\leq1}d^2\rho\: A\left(\vec{\rho}\right) \mathrm{exp} \left(\frac{2\pi i W\left(\vec{\rho}\right)}{\lambda} \right) q_{lj}\left(\vec{\rho}\right) \mathrm{exp}\left(-i\vec{k}\left(\vec{\rho}\right)\cdot \vec{r}\right) ,
\label{eq:wljdef}
\end{equation}
with $c_n$ a normalization factor, $A\left(\vec{\rho}\right)$ the amplitude, $q_{lj}\left(\vec{\rho}\right)$ the polarization vector components, defined elsewhere \cite{Stallinga2015}, and $W\left(\vec{\rho}\right)$ the aberration function, which is the sum of a contribution from the microscope system and a contribution from the SLM. The wavevector $\vec{k}(\vec{\rho})$ is a function of the pupil coordinates:
\begin{equation}
\vec{k}\left(\vec{\rho}\right) = \frac{2 \pi}{\lambda} \left(\mathrm{NA} \rho_x,\mathrm{NA} \rho_y,\sqrt{n^2 - \mathrm{NA}^{2}{\vec{\rho}}^2}\right) ,
\end{equation}
with $n$ the refractive index of the medium and NA the numerical aperture of the objective lens. The expected photon count $\mu_k$ for a given camera pixel is given by the integration of the PSF over the pixel area $D_k$ with size $a\times a$: 
\begin{equation}
\mu_k = \int_{D_k} dx dy\: H_{\mathrm{bead}}\left(\vec{r}-\vec{r}_0\right) ,
\end{equation}
with $r_0$ the position of the emitter. The derivatives with respect to the fit parameters $(\theta = x,y,z,\lambda)$ which are needed for the fitting routine are similar to \cite{Smith2016} but now involve a convolution:
\begin{equation}
\frac{\partial \mu_k}{\partial \theta_m} = \frac{2N}{3}\sum_{l=x,y}\sum_{j=x,y,z} \int_{D_k} dx dy\: \mathrm{Re}\left[w_{lj}\left(\vec{r}-\vec{r}_0\right)^{*}\frac{\partial w_{lj}\left(\vec{r}-\vec{r}_0\right)}{\partial \theta_m}\right] \otimes \bigcirc\left(\vec{r}-\vec{r}_0\right) .
\end{equation}
The derivatives of $w_{lj}$ with respect to the fit parameters remain the same:
\begin{eqnarray}
\nonumber
\frac{\partial w_{lj}\left(\vec{r}-\vec{r}_0\right)}{\partial \vec{r}_0} &=& \frac{i}{c_n }\int_{|\vec{\rho}|\leq1} d^2\rho\: A\left(\vec{\rho}\right) \mathrm{exp}\left(\frac{2 \pi i W\left(\vec{\rho}\right)}{\lambda} \right) q_{lj} \vec{k}\left(\vec{\rho}\right)\\ &&\mathrm{exp}\left(-i\vec{k}\left(\vec{\rho}\right)\cdot \left(\vec{r}-\vec{r}_0\right)\right) ,
\end{eqnarray}
and:
\begin{eqnarray}
\nonumber
\frac{\partial w_{lj}\left(\vec{r}-\vec{r}_0\right)}{\partial \lambda} &=& -\frac{2 \pi i}{c_n \lambda^2 }\int_{|\vec{\rho}|\leq1} d^2\rho\: A\left(\vec{\rho}\right) \mathrm{exp}\left(\frac{2 \pi i W\left(\vec{\rho}\right)}{\lambda} \right) q_{lj} W\left(\vec{\rho}\right) \\
& &\mathrm{exp}\left(-i\vec{k}\left(\vec{\rho}\right)\cdot\left(\vec{r}-\vec{r}_0\right)\right) +\frac{1}{\lambda}\left(\vec{r}-\vec{r}_0\right)\cdot\frac{\partial w_{kj}\left(\vec{r}-\vec{r}_0\right)}{\partial \vec{r}_0}.
\end{eqnarray}

\subsection{Aberration retrieval from a through focus PSF scan}
Aberration retrieval is done by adapting the vectorial fitting routine to estimate the aberration coefficients from a through-focus image stack of a fluorescent bead. The aberration function is expressed as a linear sum of root mean square (rms) normalized Zernike polynomials $\bar{Z}_n^m(\vec{\rho})$:
\begin{equation}
w_{lj}\left(\vec{r}\right) = \frac{1}{c_n} \int_{|\vec{\rho}|\leq1}d^2\rho\: A\left(\vec{\rho}\right) \mathrm{exp} \left(\frac{2\pi i \sum_{n,m} A_n^m \bar{Z}_n^m\left(\vec{\rho}\right)}{\lambda} \right) q_{lj}\left(\vec{\rho}\right)\mathrm{exp}\left(-i\vec{k}\left(\vec{\rho}\right)\cdot\vec{r}\right) ,
\end{equation}
where the appearing Zernike coefficients $A_n^m$ are the fit parameters. The derivatives of the electric field components with respect to the Zernike coefficients needed for the MLE fitting routine are found to be:
\begin{eqnarray}
\nonumber
\frac{\partial w_{lj}\left(\vec{r}\right)}{\partial A_k^i} &=& \frac{2 \pi i}{c_n \lambda}\int_{|\vec{\rho}|\leq1} d^2\rho\: A\left(\vec{\rho}\right) \mathrm{exp}\left(\frac{2 \pi i}{\lambda}\sum_{n,m} A_n^m \bar{Z}_n^m\left(\vec{\rho}\right)\right) \\
&&q_{lj} \bar{Z}_k^i\left(\vec{\rho}\right)\mathrm{exp}\left(-i\vec{k}\left(\vec{\rho}\right)\cdot\vec{r}\right) .
\end{eqnarray}

\section{Experimental setup and methods}
\subsection{Experimental setup}
\begin{figure}
\centering
\includegraphics[width=0.75\linewidth]{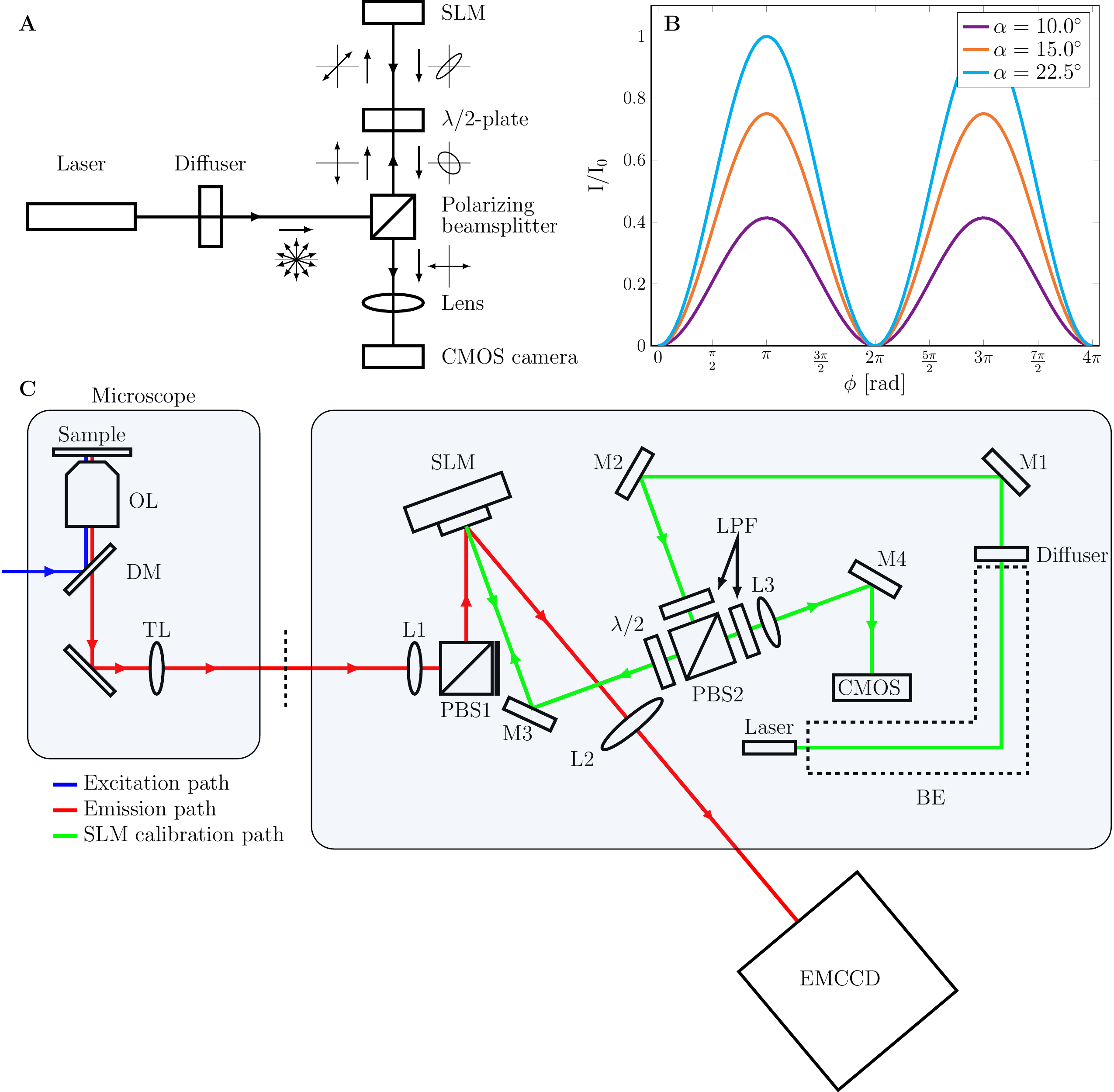}
\caption{A) Schematic drawing of the SLM calibration branch and the polarization transfer through the light path. Additional linear polarization filters are not drawn as they are aligned with the polarizing beamsplitter. B) The intensity response at the camera as a function of the phase retardation of the SLM for different orientations $\alpha$ of $\lambda/2$-plate. C) Schematic overview of the optical setup. A relay system with SLM is added to the emission path of the microscope (red) and a separate SLM calibration path (green) is incorporated into the emission relay system. This allows for SLM calibration between experiments.  BE: beam expander, DM: dichroic mirror, L: lens, LPF: linear polarizing filter, M: mirror. OL: objective lens, PBS: polarizing beam splitter, TL: Tube lens.  }
\label{fig:Opticalsystem}
\end{figure}

The setup, shown in Fig.~\ref{fig:Opticalsystem}, consists of a Nikon Ti-E microscope with a TIRF APO objective lens (NA~=~1.49, M~=~100), a 200~mm tube lens, a relay system with an SLM is built on one of the exit ports of the microscope. The relay system consists of two achromatic lenses ($f_1$~=~100 mm, Thorlabs AC254-100-A and $f_2$~=~200 mm, Thorlabs AC508-200-A), a nematic Liquid Crystal On Silicon (LCOS) SLM (Meadowlark, XY-series, 512x512 pixels, pixel size~=15~$\mu$m, design wavelength~=~532 nm) and a polarizing beam splitter to filter the $x$-polarized light which is not modulated by the SLM. The first achromatic lens relays the light on the SLM in a beam with a diameter of 3~mm and the second relay lens ensures Nyquist sampling of the fluorescent objects at the EMCCD (Andor iXon Ultra - X987, 512x512 pixels, pixel size~=~16~$\mu$m, backprojected to object space 80~nm). The microscope is equipped with a set of lasers with wavelengths 405~nm (Coherent Cube), 488~nm (Coherent Sapphire), 561~nm (Coherent Sapphire), and 642~nm (MPB Communications). Either a dichroic filter set for the green (Ex: Semrock FF01-460/60-25, Di: Semrock Di02-R532-25X36, Em: Semrock FF01-545/55-25) or a quadband dichroic filter set (Chroma - TRF89902) is used.

This standard setup is augmented with a novel, second light path for calibration of the SLM. This SLM calibration light path is designed for measuring the retardation difference between the $x$ and $y$-polarized light incident on the SLM and consists of a laser (Thorlabs - CPS532, wavelength 532~nm, 0.9~mW) which illuminates the SLM via a beam expander, a polarizing beam splitter and a $\lambda/2$ wave-plate. The reflected and polarization-modulated light passes the $\lambda/2$ wave-plate and polarizing beam splitter again and is imaged onto a CMOS camera (Thorlabs - DCC1545M, 1280x1024 pixels, pixel size = 5.2 $\mu$m). A rotating diffuser is added to reduce speckle and two linear polarizing filters, which are aligned with the polarization axes of the polarizing beamsplitter, are added to reduce internal reflections. The intensity image captured by the CMOS camera is mapped to the intensity pertaining to specific SLM pixels $I_\mathrm{pxl}$. The polarization transfer through the $\lambda/2$ wave-plate and the SLM for the calibration light path is described by the Jones-matrix:
\begin{eqnarray}
\nonumber
{\cal J} &=& {\cal J}_{\lambda/2}^{T} {\cal J}_{SLM}{\cal J}_{\lambda/2}\\
\nonumber 
&=& \left[\begin{array}{cc}
\cos(2\alpha) &-\sin(2\alpha) \\
\sin(2\alpha) & \cos(2\alpha)
\end{array}\right]
\left[\begin{array}{cc}
\exp\left(i\phi_\mathrm{pxl}\right) &0 \\
0 & 1
\end{array}\right]
\left[\begin{array}{cc}
\cos(2\alpha) &\sin(2\alpha) \\
-\sin(2\alpha) & \cos(2\alpha)
\end{array}\right]\\
\nonumber
&=& \left[\begin{array}{cc}
\cos^{2}(2\alpha)\exp\left(i\phi_\mathrm{pxl}\right) + \sin^{2}(2\alpha) &\sin(2\alpha)\cos(2\alpha)\left(\exp\left(i\phi_\mathrm{pxl}\right)-1\right) \\
\sin(2\alpha)\cos(2\alpha)\left(\exp\left(i\phi_\mathrm{pxl}\right)-1\right) & \cos^{2}(2\alpha) + \sin^{2}(2\alpha)\exp\left(i\phi_\mathrm{pxl}\right)
\end{array}\right],\\
\end{eqnarray}
with $\alpha$ the angle of the $\lambda/2$ wave-plate and $\phi_\mathrm{pxl}$ the retardance induced by an SLM pixel, and gives rise to an intensity:
\begin{equation}
\frac{I_\mathrm{pxl}}{I_0} = \left|{\cal J}_{12}\right|^{2} = 4\sin^2\left(\frac{\phi_\mathrm{pxl}}{2}\right) \cos^2(2\alpha)\sin^2(2\alpha) ,
\label{formula:Ipxl}
\end{equation}
with $I_0$ the incident illumination intensity. The waveplate angle $\alpha$ is set to a small value, around 5~deg, in order to have a small maximum transmission. This is required for having a relatively long integration time, about an order of magnitude more than the rotation period of the rotating diffuser, for enabling good speckle reduction.  

\subsection{SLM calibration}
In order to measure the modulation of a certain SLM pixel the mapping from the SLM onto the camera of the calibration path is needed. This mapping is obtained by applying a checkerboard pattern with increasing voltages to the SLM. The difference between the average captured image and the image when no voltage is applied is used as input for a corner detection algorithm (\textit{findcheckerboard} from Matlab - \textit{Mathworks}) to find the corner points. An affine transformation is fitted to these points and used to find the CMOS pixels corresponding to each SLM pixel. 

\begin{figure}
\centering
\includegraphics[width=0.95\linewidth]{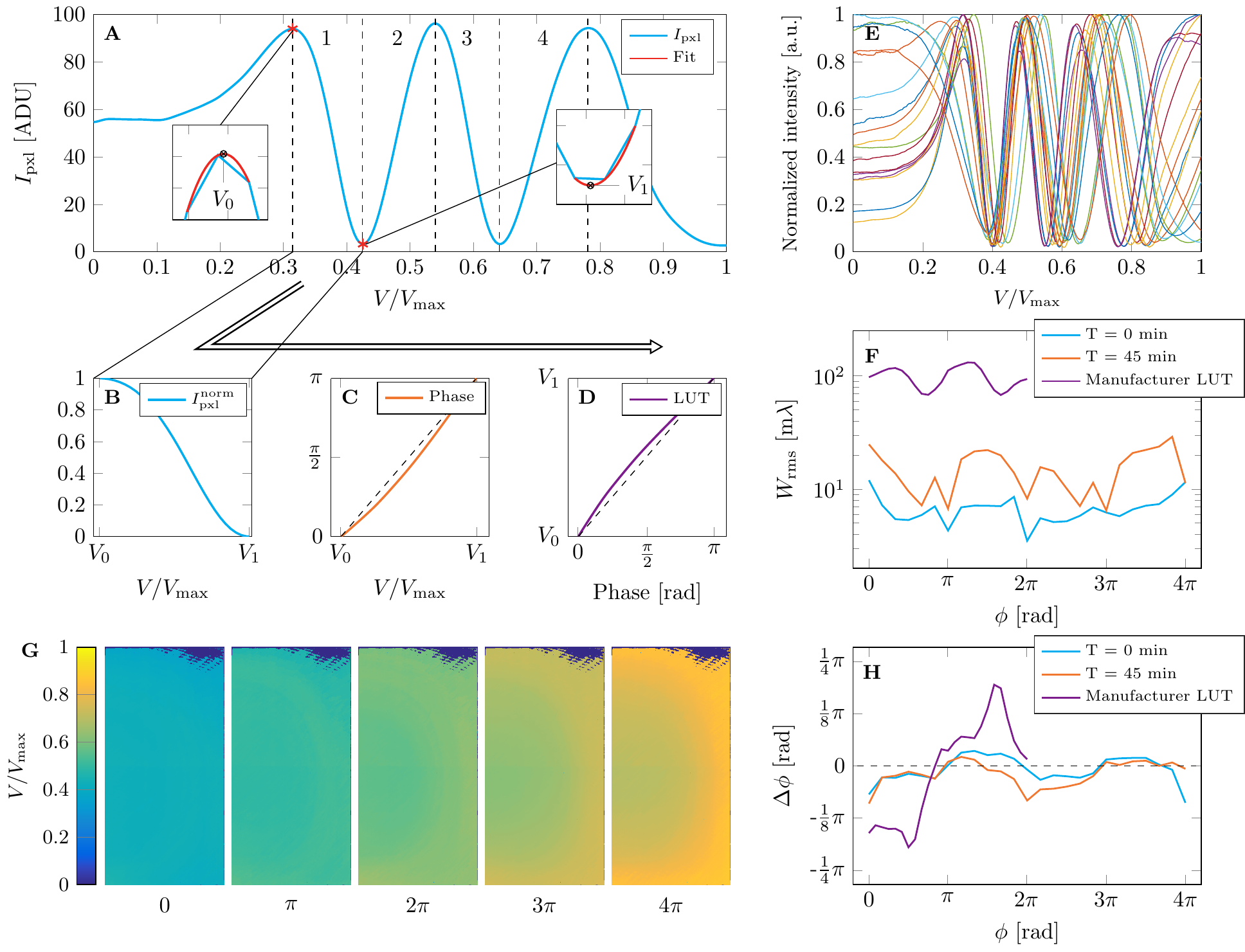}
\caption{SLM calibration procedure. A) The measured intensity response for a single SLM pixel as a function of applied voltage. Every extremum corresponds to a phase change equal to an integer multiple of $\pi$ and a second order polynomial is fitted to increase the precision in finding the extrema. The intensity is segmented into four parts which are scaled to [0 1]. This normalized intensity (B) is converted into phase (C) and inverted to create the LUT for that particular voltage segment and pixel (D). E) The normalized intensity response for 20 randomly selected SLM pixels, showing the pixel-to-pixel variations. F) The measured root mean square error of the wavefront as a function of the phase with calibration LUTs immediately after calibration, after 45 minutes, and the LUT provided by the manufacturer. G) The LUTs of the part of the SLM used in the imaging light path for different constant phases. Dark spots indicate pixels without 3 maxima. H) The difference between the measured average phase and the intended phase as a function of the intended phase. }
\label{fig:SLMcalibration}
\end{figure}

The calibration procedure for an SLM pixel is graphically explained in Fig.~\ref{fig:SLMcalibration}. First, the intensity response as function of applied voltage is measured in 256 steps, giving rise to a sequence of minima and maxima, which correspond to a retardation of $\pi$ or $2\pi$. All pixels inside the illuminated SLM plane appear to have three maxima, implying a total phase modulation of 4$\pi$ or 1094~nm. The voltages for which these extrema occur are found by fitting parabolae to the three points near the extrema, which increases the precision and fully utilizes the 16 bit control of the SLM. The intensity is then divided into four segments which are scaled to [0 1] and converted to phase using the inverse of Eq.~(\ref{formula:Ipxl}) over these segments. The phase response is used to construct an individual Look Up Table (LUT) for each SLM pixel, compensating the non-uniformity of the SLM. The LUT-parameters vary smoothly over the SLM and correspond roughly with the Fabry-Perot fringes visible by eye, indicating that the differences in phase response are due to variations of the thickness of the liquid crystal layer. Additional pixel-to-pixel variations may arise from pixel-to-pixel variations in the underlying silicon switching circuitry. The complete calibration takes about 5 minutes (3 minutes scanning and 2 minutes computing time on a quadcore 3.3 GHz i7 processor), but can in principle be optimized to run faster. 

The calibration is verified by applying a uniform retardation profile over the complete calibration range to measure the root mean square error of the reflected wavefront and compare the average intensity to the ideal intensity response. This verification is performed immediately after calibration and also after 45 minutes, and subsequently compared to the calibration provided by the manufacturer, which is only for 2$\pi$ modulation, see Fig.~\ref{fig:SLMcalibration}(F). The rms error of the wavefront with the manufacturer's LUT is around 100 m$\lambda$ and within specification, but too high for the stated goal of our research: to achieve wavefront control well below the diffraction limit in order to optimally apply complex engineered PSFs in single emitter localization. The individual pixel calibration method reduces the wavefront error by an order of magnitude to around 10 m$\lambda$ immediately after the calibration and to 20 m$\lambda$ after 45 minutes. The deterioration of performance over time is attributed to temperature fluctuations and the associated mechanical drift of the different optical components. We conclude that there is about a half hour window of opportunity to conduct experiments with a precision of the wavefront control within 20 m$\lambda$ rms wavefront aberration. All subsequent experiments have been done within this time frame. We mention that dedicated temperature control of the setup can extend the total time with high-precision wavefront control.

\subsection{Axial and lateral alignment of the SLM}
\begin{figure}
\centering
\includegraphics[width=\linewidth]{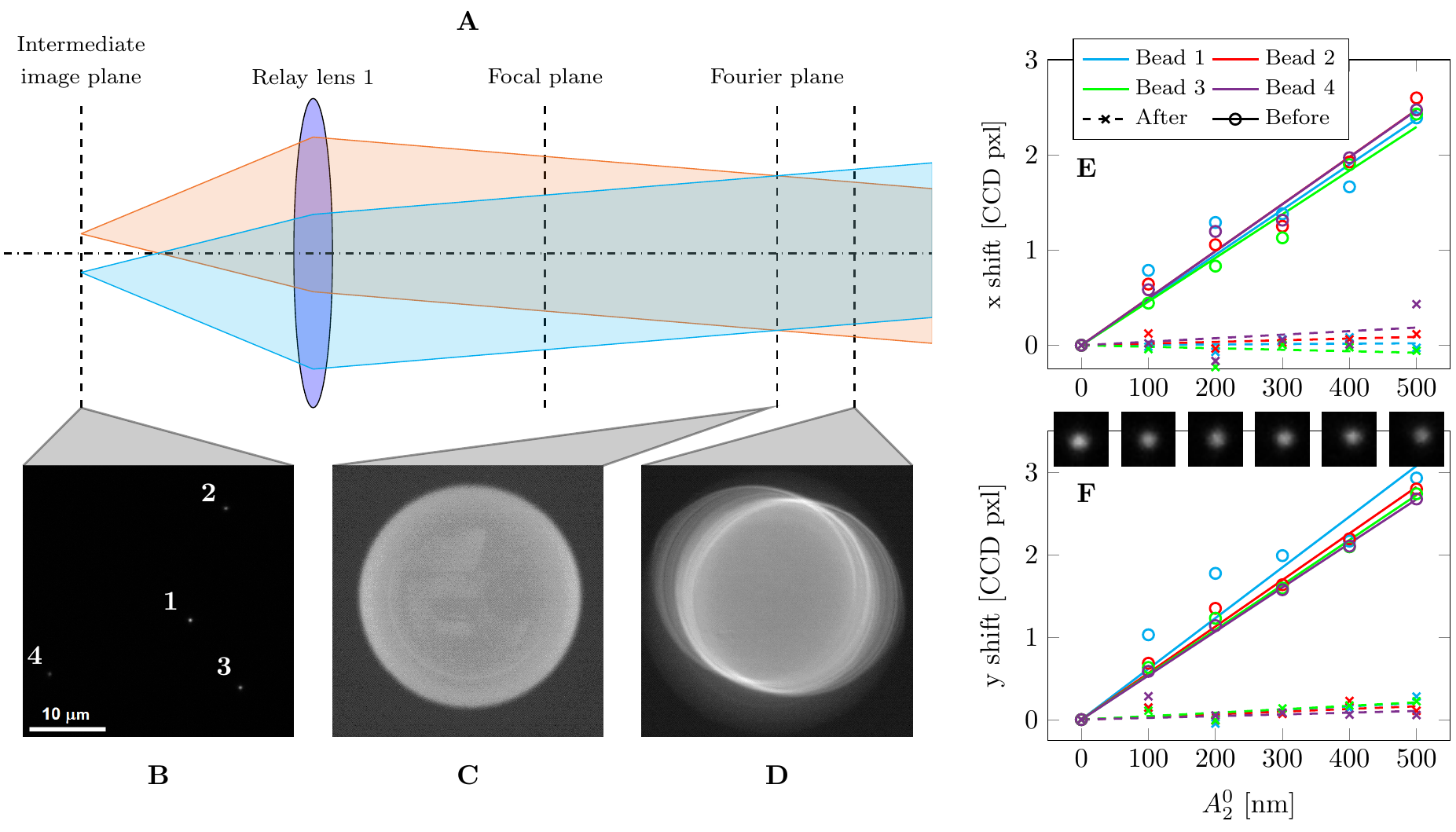}
\caption{SLM alignment principle. A) The beams collimated by the first relay lens originating from beads in the FOV (B, scalebar indicates 10~$\mu$m) will be incident on the SLM at different positions if the SLM is not aligned properly with the Fourier plane (which not necessarily coincides with the back focal plane of the relay lens). To illustrate this the SLM is temporarily replaced by a camera prior to further alignment steps. The acquired images of the beams at the correct Fourier plane or at a plane axially misaligned by approximately 5~mm are shown in C and D, respectively. The different visible circles correspond to the emission beams of the beads located at different positions in the FOV. E,F) The beads in the FOV shift laterally when a defocus aberration profile is applied by the SLM. This shift is a measure for the lateral misalignment of the aberration profile on the SLM with respect to the optical axis of the imaging system. After alignment the beads no longer shift laterally when the defocus is applied. }
\label{fig:SLMalignment}
\end{figure}

The position of the Optical Axis (OA) of the emission path on the SLM is needed in order to project the phase profile at the correct position on the SLM. Furthermore the SLM should be aligned with the plane conjugate to the pupil plane of the objective lens, the Fourier plane, to ensure that every point inside the Field Of View (FOV) has the same phase modulation. A procedure is used to directly estimate this alignment by imaging a set of beads in the FOV and applying a defocus modulation, without the need for additional optical components. For a bead in the center of the FOV the chief ray of emitted light beam aligns with the OA and intersects the SLM plane at a position ($\rho_x^{\mathrm{OA}}$,$\rho_y^{\mathrm{OA}}$) with respect to the coordinate frame that has its origin at the center of the SLM area designated for use in the aberration correction and PSF engineering. Here $\rho_x^{\mathrm{OA}}$ and $\rho_y^{\mathrm{OA}}$ are dimensionless pupil coordinates (real space coordinates normalized by the pupil radius). If the SLM introduces defocus then this parabolic aberration profile will be decentered w.r.t. the OA:
\begin{eqnarray}
\nonumber
W &=& A_2^0\left[2\left(\left(\rho_x-\rho_x^{\mathrm{OA}}\right)^2 + \left(\rho_y-\rho_y^{\mathrm{OA}}\right)^2\right)-1\right]\\
&=& A_2^0\left[2\left(\rho_x^2+\rho_y^2\right)^2-1\right] - 4A_2^0 \left(\rho_x^{\mathrm{OA}}\rho_x+\rho_y^{\mathrm{OA}}\rho_y\right) + A_2^0\left({\rho_x^{\mathrm{OA}}}^{2}+{\rho_y^{\mathrm{OA}}}^{2}\right),
\label{eq:shiftdefocus} 
\end{eqnarray}
where $A_2^0$ is the Zernike fringe (not rms normalized) coefficient for defocus, and where $W$ has length units. Apparently, not only defocus is introduced to the overall imaging system but also tip and tilt. As a consequence, the experimental PSF obtained from bead images will shift laterally, where the shift varies linearly with the applied defocus according to: 
\begin{equation}
\Delta x = 4\frac{\rho_x^{\mathrm{OA}}}{\mathrm{NA}}A_2^0, 
\end{equation}
(and similarly for the $y$-direction), as follows from comparing the tip/tilt in Eq.~(\ref{eq:shiftdefocus}) to the $\vec{k}\left(\vec{\rho}\right)\cdot \vec{r}$ term in Eq.~(\ref{eq:wljdef}). Therefore, the lateral alignment of the SLM can be estimated from the shift of a bead in the center of the FOV and the axial alignment can be estimated from the shift of the beads at the rim of the FOV. The center of these beams will intersect the SLM at a different lateral position if the SLM is not aligned with the Fourier plane as illustrated in Figs.~\ref{fig:SLMalignment}(A)-\ref{fig:SLMalignment}(D). The images of Fig.~\ref{fig:SLMalignment}(C) and~\ref{fig:SLMalignment}(D) where taken by replacing the SLM with a camera (Thorlabs - DCC1545M) and the camera alignment served as an estimate for the axial alignment of the SLM. The position of the beads is estimated from bright bead images (high signal-to-background) using a centroid weighted fit and the shift is measured for multiple defocus coefficients and then fitted with a linear line in order to reduce the effects of manual over and under focusing as shown in Figs.~\ref{fig:SLMalignment}(E)-\ref{fig:SLMalignment}(F). The procedure results in an initial lateral misalignment of the OA with (8,10)$\pm$0.3 SLM pixels, and after correction of the SLM aberration center with only (0.2,0.1)$\pm$0.3 SLM pixels (about 3~$\mu$m error), indicating that the phase profile could be aligned with the optical axis in a single iteration. The differences in shifts of the beads at the rim of the FOV of less than about 0.5~SLM pixels indicates that the SLM is aligned with the Fourier plane within approximately 100~$\mu$m.

\subsection{Correction of oblique angle of incidence at SLM}
The beam is reflected by the SLM under an oblique angle of $\theta$~=~20 degrees. The aberration function $W\left(x,y\right)$ added by the SLM is related to the aberration function $W_\mathrm{slm}\left(x',y'\right)$ under normal incidence, where $\left(x,y\right)$ are the pupil coordinates normal to the beam axis and $\left(x',y'\right)$ are the coordinates in the SLM plane, by:
\begin{equation}
W_\mathrm{slm}\left(x',y'\right) = \frac{1}{\cos\theta'}W\left(x'\cos\theta,y'\right) ,
\label{Eq:obliqueSLM}
\end{equation}
with $\theta'$ the corresponding angle of incidence inside the SLM ($\sin\theta = n \sin\theta'$, with $n$ the refractive index inside the SLM, taken to be $n=1.5$, and where the birefringence within the liquid crystal is neglected). Equation~(\ref{Eq:obliqueSLM}) describes two effects. The first is the scaling of the phase depth due to the oblique incidence, the second is the projection of the circular pupil onto the SLM plane, giving an anisotropic stretch and hence an elliptical cross-section (aspect ratio $\cos\left(20\deg\right)=0.94)$. Implementing this transformation ensures that the phase profile contributed by the SLM corresponds to the aberration function for normal incidence used in the localization algorithm. 

\subsection{Sample and fluorescence emission spectra}
\begin{figure}
	\centering
	\includegraphics[width=\linewidth]{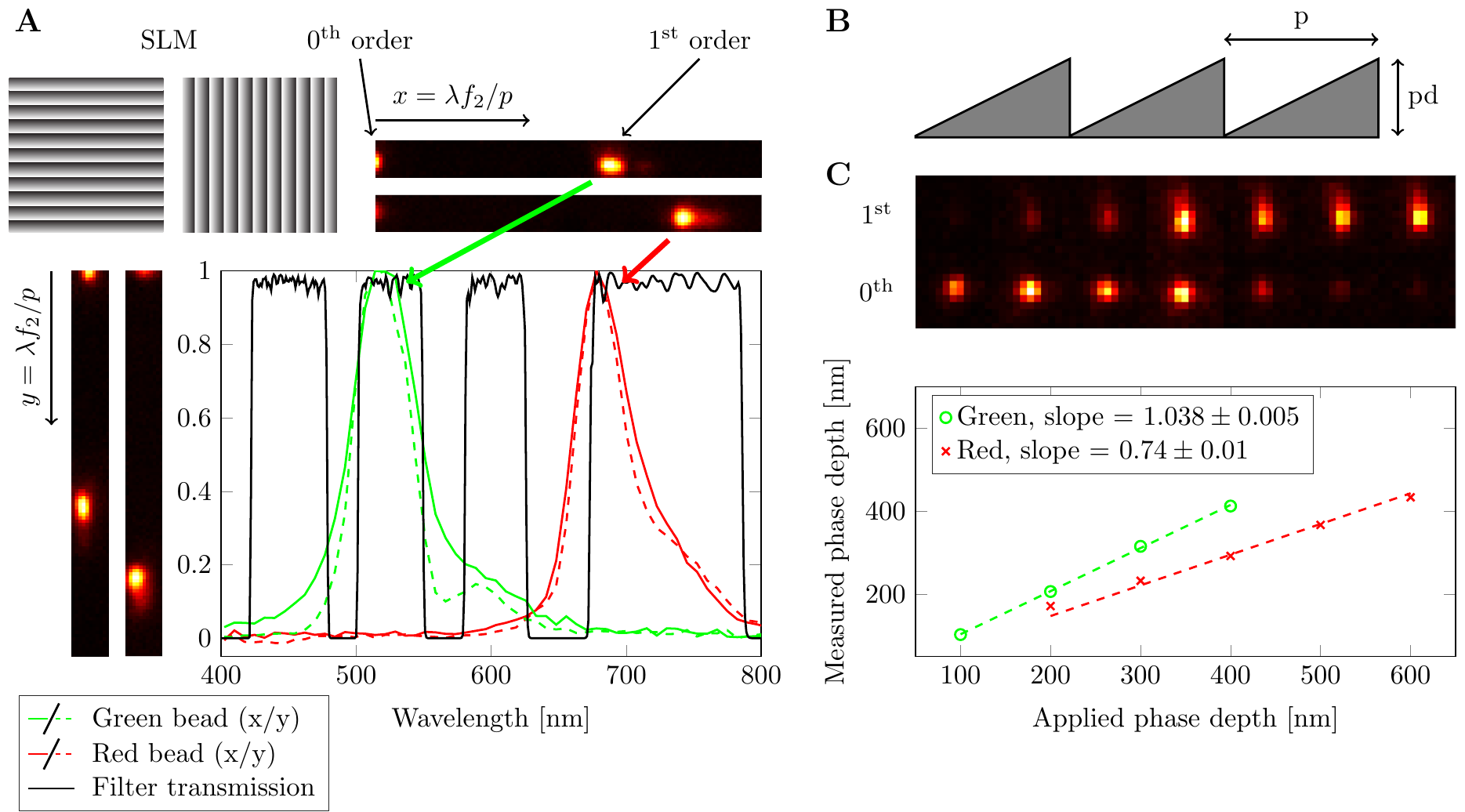}
	\caption{A) The emission spectra of the red and green beads are measured by applying a blazed grating to the SLM. The distance between the zeroth and first order on the camera is then a measure for the wavelength, the spread of the first order spot is a measure for the emission spectrum. The spectra for gratings applied in the $x$ and $y$ direction (full and dashed lines) are identical, confirming the oblique angle correction Eq.~(\ref{Eq:obliqueSLM}). B) Illustration of the blazed grating and pitch $p$ and path length step $pd$. C) The phase depth is calibrated by fitting the measured intensity ratio between the zeroth and first order with Eq.~(\ref{Eq:intensityratio}). }
	\label{fig:spectralanalyses}
\end{figure}

All proof-of-principle experiments are performed on fluorescent beads emitting in the green (Ex/Em peaks: 505/515 nm, size = 175 nm, ThermoFisher - PS-Speck) with 488~nm laser excitation and the red (Ex/Em: 660/680 nm, size = 200 nm, ThermoFisher - TetraSpeck) with 642~nm laser excitation. The weighted average emission wavelength of the green emitting bead, weighted with the product of the specified emission spectrum and the filter spectra, is 536~nm (Chroma quadband filter) or 552~nm (Semrock green filter), the weighted average emission wavelength of the red emitting bead is 692~nm. The beads are put on a cover slip and immersed in oil (n = 1.51) to guarantee the best possible refractive index matching. The emission spectra of the fluorophores are measured by introducing a blazed grating profile at the SLM (pitch~=~100~$\mu$m, maximum path length modulation $pd~=~500$~nm), see Fig.~\ref{fig:spectralanalyses}(A) for the results. The relative large pitch is sufficient to make a rough estimate of the emission spectra, but is not comparable to the spectral resolving power of a spectrometer. The spectral broadening is mainly due to the diffraction limited spot size on the camera, leading to an apparent non-zero emission in the bandstop regions of the dichroic. The spectral resolution is estimated to be on the order of 40~nm (estimate obtained from the ratio of the 0th order spot size to the distance between the 0th and 1st order times the peak emission wavelength). Blazed grating profiles are applied in the $x$ and $y$-direction (the direction of incidence is tilted in the $x$-direction at the SLM) giving rise to substantially the same emission spectrum, thereby confirming that the anisotropic stretch of the SLM aberration function described by the obliquity correction of Eq.~(\ref{Eq:obliqueSLM}) is correct. The emission peaks are found at around 520~nm and 680~nm, and the weighted average emission wavelengths are 536~nm and 693~nm, for the green and red beads, respectively, matching the bead specifications. The path length modulation of the SLM $pd$ is calibrated by applying a blazed grating with increasing phase depth. This results in diffraction orders $m=0,1,\ldots$ with amplitudes:
\begin{equation}
C_{m}=\int_{0}^{1}dt\,\exp\left(\frac{2\pi i pd t}{\lambda}\right)\exp\left(-2\pi i m t\right) = \mathrm{sinc}\left(\frac{pd}{\lambda}-m\right),
\end{equation}
with $\mathrm{sinc}\left(x\right)=\sin\left(\pi x\right)/\left(\pi x\right)$, giving an intensity ratio between the zeroth and first diffraction order:
\begin{equation}
\frac{I_0}{I_1} = \frac{\mathrm{sinc}\left(pd/\lambda\right)^2}{\mathrm{sinc}\left(pd/\lambda-1\right)^2}.
\label{Eq:intensityratio}
\end{equation}
By measuring this intensity ratio it is possible to estimate the phase depth modulated by the SLM and compare this to the expected applied phase depth modulation as shown in Fig.~\ref{fig:spectralanalyses}(C). This results in a phase depth modulation which is 3.8\% higher for the green bead and can be regarded as a small correction of the $1/\cos(\theta')$ term in the oblique angle correction. The measured phase depth in the deep red is 74\% of the phase depth expected from the SLM calibration experiments. This may be due to fringe field effects between the pixels \cite{Ronzitti2012,Persson2012,Lingel2013}, possibly in relation to the dielectric mirror and coatings that are optimized for green light, to inherent chromatic dispersion of the liquid crystal and/or to spurious reflections in the system contributing to the 0th order. Both differences in modulation are incorporated into the MLE-based localization/wavelength fitting routine by correcting the phase modulation with a factor 1.038 and 0.74 for the green and red bead, respectively. The observed variation of effective phase depth with wavelength is not incorporated into the calibration of the SLM for pixel-to-pixel variations with the additional calibration light path.

\section{Experimental results}
\subsection{Aberration retrieval}
\begin{figure}
	\centering
	\includegraphics[width=\linewidth]{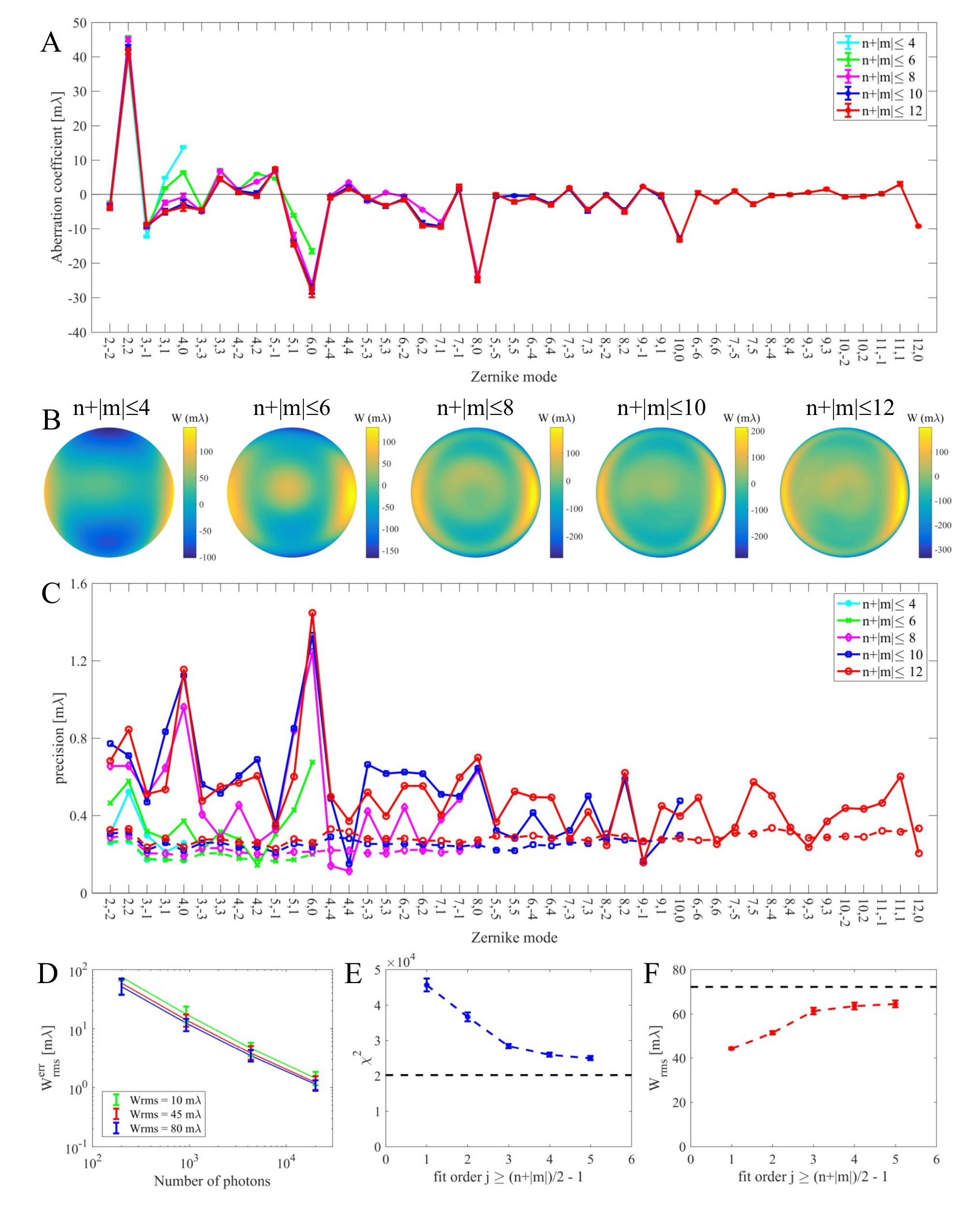}
	\caption{A) Fitted Zernike aberrations coefficients (rms values) of the non aberration corrected microscope for different sets of modes taken into account. Fits have been done for modes $n+\left| m\right|\leq 2\left(j+1\right)$ with $j=1,2,3,4,5$ the fit order. The coefficients found from higher order fits match reasonably well with the values found from the lower order fits. B) The retrieved aberrated wavefronts according to the fitted Zernike coefficients. C) The fit precision found in experiment (full lines) and according to the CRLB (dashed lines). D) Fit precision in simulation (data points) and CRLB of the fit (solid lines) as a function of photon count for different total rms aberration levels. E) The $\chi^{2}$ value of the fit as a function of fit order $j$, flattening off at a value about 20\% higher than the expected value (dashed line). F) The rms level of the fitted Zernike coefficients as a function of fit order $j$ converging to a value just below the diffraction limit (dashed line).}
	\label{fig:Zernikeorders}
\end{figure}

The aberration retrieval and subsequent correction is tested experimentally for beads emitting in the green, using the dichroic filter set for green emission. In particular, the dominant Zernike aberration modes, the fit precision, and the goodness of fit is assessed. To this end through-focus PSF stacks of 21 slices in a 2~$\mu$m range are recorded, converted to photon counts with a gain calibration procedure, and fitted on a 31$\times$31 pixel Region Of Interest (ROI), where the 3D-position of the bead, the number of signal photons, the number of background photons per pixel, and a set of Zernike aberration coefficients are used as fit parameters in the MLE fitting routine. Measurements are repeated 5~times for determining the precision. The fitted signal photon count per focal slice is around 2.4$\times 10^{4}$, the fitted background photon count around 19 photons/pixel. Figure~\ref{fig:Zernikeorders}(A) shows the retrieved Zernike aberration coefficients for fitted modes $(n,m)$ satisfying $n+\left| m\right|\leq 2\left(j+1\right)$ with $j=1,2,3,4,5$ the fit order ($j=1$ includes primary astigmatism, coma, and spherical aberration, $j=2$ includes secondary astigmatism, coma, and spherical aberration, and primary trefoil, etc.), Fig.~\ref{fig:Zernikeorders}(B) shows the corresponding retrieved wavefronts. The Zernike coefficients found at fit order $j$ match well with the ones found at a lower fit order $j-1$, indicating that there is little cross-talk between Zernike modes in the MLE fitting routine, and that there is no overfitting with large numbers of fitted aberration coefficients. Even for the case $j=5$ for which 45 modes are retrieved, there seem to be no problems with convergence (typically only about 6 iterations are needed), and reproducibility of the fit. The dominant Zernike modes appear to be primary astigmatism ($A_{2}^{2}$), secondary coma ($A_{5}^{1}$), and two higher orders spherical aberration ($A_{6}^{0}$, $A_{8}^{0}$). The correction collar of the objective lens is used to reduce the primary spherical aberration ($A_{4}^{0}$) as much as possible, but this does not seem to compensate for the higher orders of spherical aberration.

The experimental fit precision for all Zernike modes is below 1.5~m$\lambda$ and typically around 0.6~m$\lambda$, somewhat higher than the CRLB, which is around 0.3~m$\lambda$ [Fig.~\ref{fig:Zernikeorders}(C)]. The impact of photon count on the fit precision of the aberration retrieval is further assessed with a simulation study. To this end through-focus PSF stacks of 21 slices in a 2~$\mu$m range are simulated where the total rms aberration level is kept constant at 10, 45 and 80~m$\lambda$. We use $N_{\mathrm{cfg}}=100$ random instances per aberration level. The set of aberrations used consists of all Zernikes modes of radial order $n\leq 4$ (except piston, tip, tilt, defocus). Next, shot noise is added corresponding to a range of signal photon counts $N_{\mathrm{ph}}$ and a background of $b=10$ photons per camera pixel, and these sets of noisy through-focus PSFs are used as input for the MLE aberration fitter. The fit precision is quantified by the average rms error of the wavefront:
\begin{equation}
W_\mathrm{rms}^\mathrm{err} = \frac{1}{N_{\mathrm{cfg}}}\sum_{N_{\mathrm{cfg}}}\sqrt{ \sum_{n,m}({A_{n}^{m}}_\mathrm{true}-{A_{n}^{m}}_\mathrm{est})^2},
\end{equation}
and the quality of the fit is evaluated by comparing this fit precision to the CRLB. Fig.~\ref{fig:Zernikeorders}(D) shows the result, and indicates that the precision scales as $1/\sqrt{N_{\mathrm{ph}}}$, in agreement with expectations. The average residual wavefront error of the fit appears to be drop below 1~m$\lambda$ for a through-focus stack with more than $10^{4}$ signal photons, corresponding to the experimental conditions in the aberration retrieval tests.

The goodness of fit is estimated with a chi-square test. The chi-square statistic is defined as:
\begin{equation}
\chi^{2} = \sum_{k=1}^{K}\frac{\left(n_{k}-\mu_{k}\right)^{2}}{\mu_{k}},
\end{equation}
where $n_{k}$ is the measured photon count per pixel in each focal slice, and $\mu_{k}$ is the photon count expected from the fit model. Here $K$ is the total number of pixels in the fit region times the number of focal slices, i.e. the total number of statistically independent measurements. If the $n_{k}$ follow a Poissonian distribution with rates $\mu_{k}$ then the mean and variance of the statistical distribution of $\chi^{2}$ values follow as:
\begin{eqnarray}
\mathrm{mean}\left(\chi^{2}\right) &=& K,\\
\mathrm{var}\left(\chi^{2}\right) &=& 2K+\sum_{k=1}^{K}\frac{1}{\mu_{k}},
\end{eqnarray}
where the expectation values $\langle\left(n_{k}-\mu_{k}\right)^{2}\rangle=\mu_{k}$ and $\langle\left(n_{k}-\mu_{k}\right)^{4}\rangle=\mu_{k}+3\mu_{k}^{2}$ for the Poisson-distribution are used. The statistical distribution of $\chi^{2}$ values may be approximated by a normal distribution as $K\gg 1$, even though the statistics of each measured pixel is Poissonian. The goodness of fit can then be quantified by the level of confidence found by comparing the experimental $\chi^{2}$-value to the mean and standard deviation of the expected normal distribution of $\chi^{2}$-values. Figure~\ref{fig:Zernikeorders}(E) shows the measured $\chi^{2}$-values in relation to the expected value $K=21\times 31^{2}=2.0\times 10^{4}$. It appears that the $\chi^{2}$-value converges to a level about 20\% higher than the expected value, significantly more than the expected standard deviation 2$\times 10^{3}$. This shows that there are still some model errors left, probably effects of photobleaching and illumination intensity variations (it is assumed in the fit that the same level of signal and background photon count applies to each focal slice). Other effects which possibly contribute to the discrepancy are the non-zero spectral bandwidth of the collected fluorescence emission, residual scattering at the SLM, and amplitude aberrations or apodization due to e.g. variations in transmission through the objective lens and the dichroics with pupil position \cite{Liu2013}.

\subsection{Aberration correction}
\begin{figure}
	\centering
	\includegraphics[width=\linewidth]{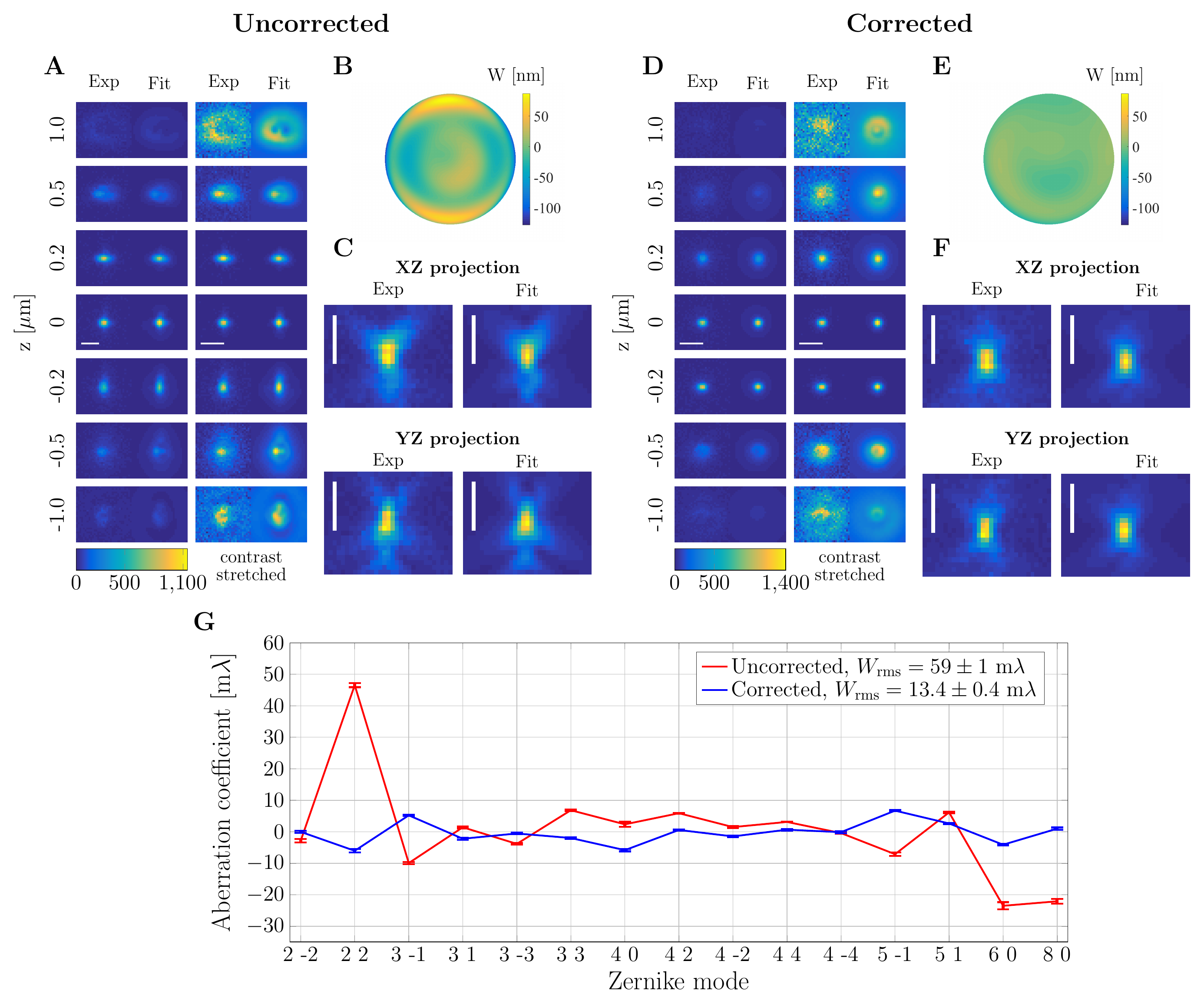}
	\caption{A-F) Measured and fitted through-focus PSF of beads emitting in the green (range: [-1, 1] $\mu$m, 21 steps) before and after aberration correction, see also \textcolor{blue}{Visualisation~1} and \textcolor{blue}{Visualisation~2}. The left columns of A and D show measured and fitted focal slices, the right columns of A and D show the same images contrast stretched with the same scale for each exp/fit pair for visibility. Scalebar indicates 1~$\mu$m. G) Fitted Zernike coefficients (rms values) before and after aberration correction showing a reduction in the $W_\mathrm{rms}$ of the model PSFs that best fit the measurements from $59\pm 1$~m$\lambda$ to $13.4\pm 0.4$~m$\lambda$.}
	\label{fig:AberrationCorrection}
\end{figure}

The overall rms wavefront error, as determined from the fitted Zernike coefficients converges to a value around 65~m$\lambda$, just below the diffraction limit, with increasing number of Zernike modes [Fig.~\ref{fig:Zernikeorders}(F)], proving that aberration correction is needed for successful application of complex engineered PSFs. There is a trade-off between the number and type of Zernike modes that can be corrected and the peak-valley value of the wavefront that is needed for the desired engineered PSFs, in view of the limited phase dynamic range of the SLM. In the experiments on engineered PSFs we take all Zernike modes with radial order $n\leq 4$ into account, similar to previous studies in the literature \cite{Roider2014,Petrov2017}. In order to evaluate the best possible performance of the aberration correction we also include second order coma ($A_{5}^{1}$ and $A_{5}^{-1}$), second order spherical aberration ($A_{6}^{0}$), and third order spherical aberration ($A_{8}^{0}$). Figure~\ref{fig:AberrationCorrection} and \textcolor{blue}{Visualisation~1} and \textcolor{blue}{Visualisation~2} show the experimental and fitted through-focus PSF without and with aberration correction. The agreement between measured and fitted PSFs is quite well, especially in the ~1~$\mu$m range around focus. It appears that the experimental through-focus PSF is more rotationally symmetric after correction, proving that asymmetry inducing aberrations are reduced. In addition, the asymmetry between spot shapes above and below focus is greatly reduced after correction, indicating that spherical aberration is largely eliminated. The wavefront error estimated with the aberration retrieval algorithm reduced from $59\pm 1$~m$\lambda$ to $13.4\pm 0.4$~m$\lambda$ [Fig.~\ref{fig:AberrationCorrection}(G)]. This residual error is close to the calibration precision of the SLM, indicating that the level of aberration correction is limited by the precision of the SLM control. 

\begin{figure}
	\centering
	\includegraphics[width=0.7\linewidth]{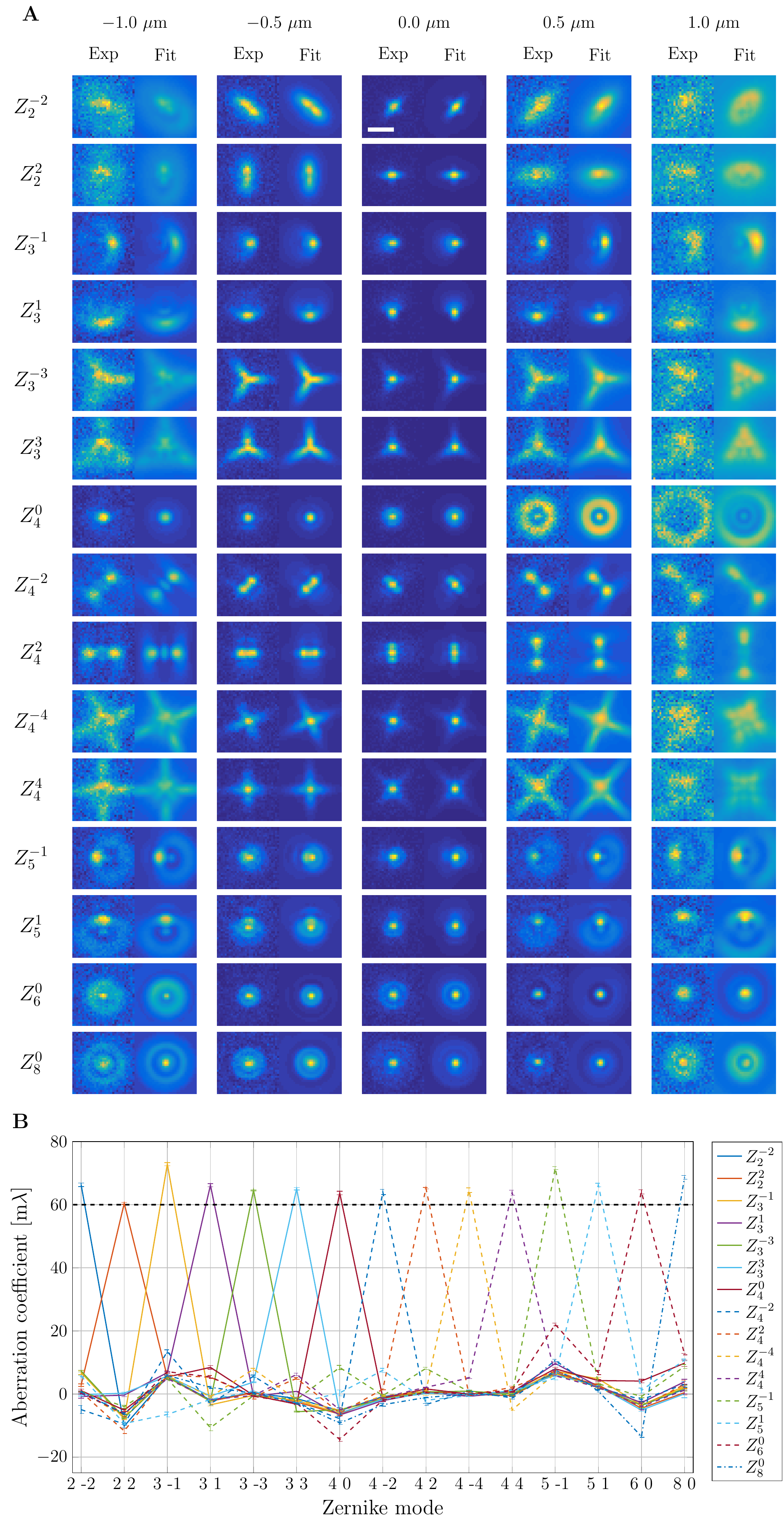}
	\caption{A) Through-focus stacks of green emitting beads deliberately aberrated by single Zernike modes with coefficients $A_n^m = 60$~m$\lambda$ and the corresponding theoretical through-focus stacks using the aberration coefficients found from the aberration fitting routine (through-focus range: [-1, 1] $\mu$m, 21 steps, estimated photon count was around 2.2$\times 10^{4}$ signal and 32 background photons). Scalebar indicates 1~$\mu$m and all exp/fit image pairs are contrast stretched with the same scale. B) The aberration retrieval appears to be mode specific, and estimates the Zernike coefficients as $67\pm 4$~m$\lambda$, averaged over the 15 displayed Zernike modes.}
	\label{fig:ZernikePSF}
\end{figure}

A final test of the aberration retrieval and correction procedure is performed by deliberately adding single Zernike modes ($A_{n}^{m} = 60~$m$\lambda$) on top of the corrected wavefront and subsequently feeding the aberrated through-focus stack to the aberration fitting routine, see Fig.~\ref{fig:ZernikePSF} for the results. The fitting routine correctly retrieves each aberration with an estimated value of $67\pm4$~m$\lambda$, averaged over the 15 displayed aberrations, somewhat higher than the expected 60~m$\lambda$. All other retrieved aberrations remain at the level of 20~m$\lambda$ or less, pointing to the specificity of the aberration retrieval and correction procedure. In particular, there is little crosstalk from added aberrations of radial order $n$ to retrieved aberrations of order $n-1$, which confirms the correct lateral alignment of the SLM phase profile with respect to the optical axis.

\subsection{Analysis of engineered PSFs}
Proof-of-principle experiments of engineered PSFs have been performed with green and red emitting beads, using the quadband dichroic filter set. Aberrations are corrected only up to radial order $n\leq 4$ in order to save phase dynamical range on the SLM for the engineered PSFs. Primary spherical aberration is compensated by the correction collar of the objective lens. Three different designs have been tested. The first is a binary grating splitting the spot into two $\pm 1$st diffraction orders, where the grating zones are curved to induce astigmatism to the two orders (grating zone shape described by Zernike coefficients $A_{1,\mathrm{zone}}^{-1} = 0.8\lambda_0$ and $A_{2,\mathrm{zone}}^{-2} = 0.15\lambda_0$, as in \cite{Smith2016}). The nominal wavelength $\lambda_0$ is equal to 520~nm for the green emitting beads and 690~nm for the red emitting beads.The second engineered PSF is a blazed grating for splitting the spot into a 0th and +1st order (grating zones curved to induce astigmatism, the shape is described by Zernike coefficients $A_{1,\mathrm{zone}}^{-1} = 1.4\lambda_0$ and $A_{2,\mathrm{zone}}^{-2} = 0.3\lambda$) and an overall continuous astigmatic aberration profile with Zernike coefficient $A_{2,\mathrm{overall}}^{-2} = -0.15\lambda_0$. The third engineered PSF is a double helix configuration (annular design~\cite{Prasad2013,Roider2014} with four rings and exponent $\alpha$ = 1/2, and phase depth = $\lambda_0$). These design parameters are set by balancing the achievable precision with the axial range and with the footprint of the spot on the detector, but can in theory be improved by e.g. incorporating higher order astigmatism in case of the astigmatic profiles (mimicking the saddle-point or tetrapod PSF \cite{Shechtman2014}). Through-focus image stacks of these engineered PSFs are recorded for different signal photon counts while keeping the background constant using the camera frame-time and the trans-illumination unit for conventional brightfield microscopy for providing extra background photons at shorter frame times as tuning parameters, and subsequently fitted using MLE and the vectorial PSF model. 

\begin{figure}
	\centering
	\includegraphics[width=\linewidth]{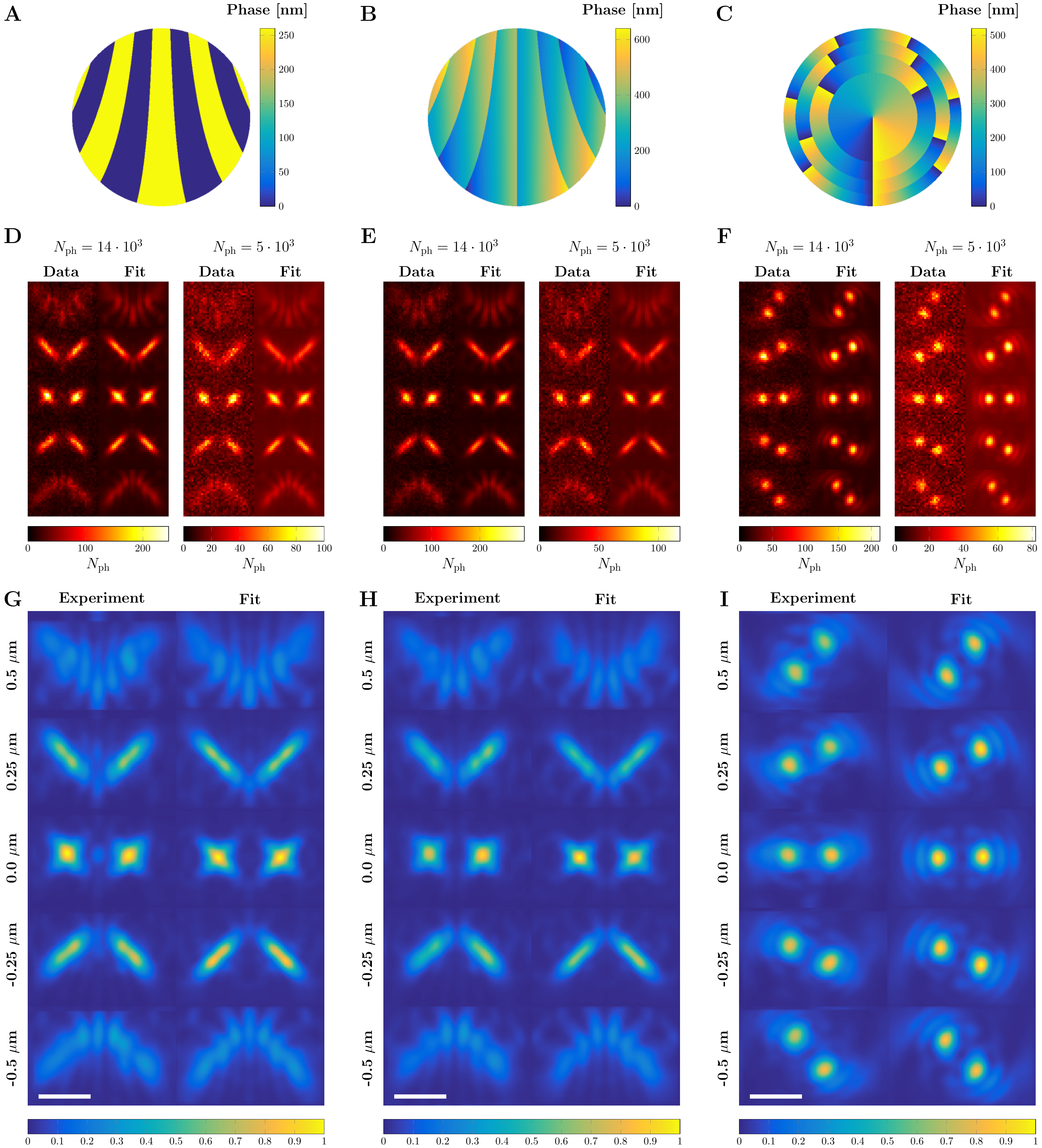}
	\caption{PSF comparison between the measured PSF and fit with the vectorial PSF model. The phase mask of the binary astigmatic PSF (A), the blazed astigmatic PSF (B) and the Double Helix (C) alongside example acquisitions and fits for two different photon counts (D-F). G-I) The average measured PSF is compiled from the signal carried by approximately $10^8$ photons by upsampling (3$\times$) and overlaying all acquired spots. The scale bars indicate 1~$\mu$m.}
	\label{fig:psfcomparison}
\end{figure}

Figure~\ref{fig:psfcomparison} shows the phase profile for the three engineered PSF designs and examples of measured spots and the corresponding fits for two estimated signal photon counts ($5\cdot10^3$ and $14\cdot10^4$). The overall experimental PSF, obtained by adding all recorded spots after first upsampling 3$\times$ and subsequently shifting with the fitted $xy$-position of the bead, is shown as well, along with the prediction of the vectorial PSF model. In this way the experimental PSF is built up from the cumulative signal of $\sim 10^8$ photons. The agreement between experiment and the theoretical vectorial PSF is generally excellent, even the fringe structures at the largest defocus values match very well. The remaining discrepancies, mainly a slight broadening of the spots, is attributed to the non-zero spectral width of the light incident on the camera, due to the width of the emission spectrum and the width of the bandpass regions of the quadband dichroic. There is also a small asymmetry in the fringe structure, which is probably caused by the residual higher order spherical aberrations in the optical system. 

\begin{figure}[tb]
	\centering
	\includegraphics[width=\linewidth]{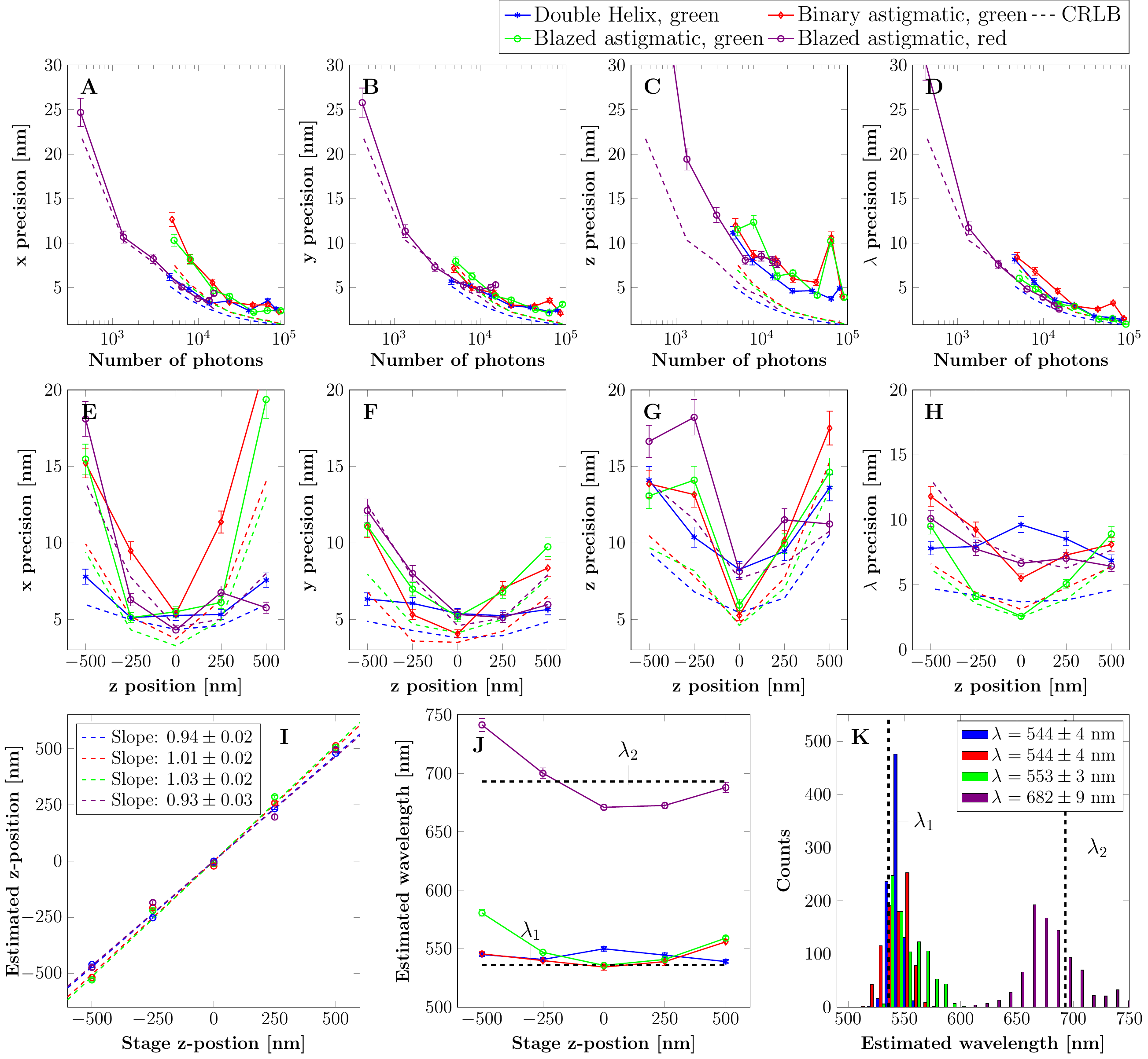}
	\caption{A-D) Achieved localization precision for the fit parameters $x/y/z/\lambda$ as a function of photon count estimated from the fit. E-H) Localization precision over the axial range for an estimated photon count around of $5\times 10^{3}$ (green bead) and $3\times 10^{3}$ (red bead). I) Average estimated $z$-position as a function of stage $z$-position for all acquisitions. J) Average estimated wavelength as a function of stage $z$-position for all acquisitions, giving values close to the weighted average emission wavelengths $\lambda_{1}=536$~nm and $\lambda_{2}=693$~nm. K) Histogram of estimated wavelength values for all acquisitions.}
	\label{fig:crlbplot}
\end{figure}

The achieved precision in $x/y/z/\lambda$ for the three engineered PSFs in focus as a function of signal photon count are shown in Figs.~\ref{fig:crlbplot}(A)-\ref{fig:crlbplot}(D), and for one signal level as a function of the axial position in Figs.~\ref{fig:crlbplot}(E)-\ref{fig:crlbplot}(H). The localization precision is obtained by fitting 25 acquisitions of a single bead at each $z$-position. The precisions follow the CRLB as a function of signal photon count and axial position, with the exception of the $z$-precision, which is somewhat worse than the CRLB. This is attributed to the higher orders of spherical aberration, which are not corrected for these datasets. The localization precision values appear to level off at values around $x/y/z = 2/2/5$~nm for high photon counts, probably due to effects of drift. The overall performance of the three engineered PSFs concerning precision appears to be quite similar. Figure~\ref{fig:crlbplot}(F) shows the average fitted $z$-position as a function of stage $z$-position. Both astigmatic engineered PSFs have a linear response with a fitted slope of 1.03$\pm$0.02 (blazed) and 1.01$\pm$0.02 (binary) in the green, but the double helix PSF and the blazed astigmatic PSF in the red underestimate the $z$-position slightly with a slope of 0.94$\pm$0.02 and 0.93$\pm$0.03, respectively. This bias may also be due to uncorrected higher order spherical aberration. The estimated wavelengths shown in Figs.~\ref{fig:crlbplot}(J)-\ref{fig:crlbplot}(K) are close to the measured weighted average emission wavelengths and resolve the green and red bead excellently. There seems to be a small overestimation of the wavelength for the green bead, which is constant over the entire axial range for the binary astigmatic PSF and the double helix PSF, but not for the blazed astigmatic PSF. The accuracy of the $z$-position is most likely affected by these small over and underestimations of the wavelength, as these parameters are closely coupled in the fitting routine. 

\section{Conclusion} 
In summary, we have shown how a dedicated calibration protocol for high-NA fluorescence microscopes equipped with an adaptive optical element such as an SLM can reduce the aberrations to around 20~m$\lambda$ rms wavefront aberration. This high level of wavefront control enables single emitter localization with complex engineered PSFs, where the full vectorial PSF model is used in the fitting routine. A key ingredient in the calibration protocol is a separate SLM calibration light path for suppressing the effect of the non-uniformity of the SLM. In these well-controlled experimental circumstances the experimental PSFs conform very well to the predictions of the vectorial PSF model. The method is further tested with proof-of-principle experiments for fitting the 3D-position and the emission wavelength of single emitters. The precision in $x/y/z/\lambda$ is similar to the CRLB in all four fit parameters, and a high accuracy in $z$ and $\lambda$ is obtained, without an experimental reference PSF or LUT.

An open issue is the residual model mismatch, as revealed by the $\chi^{2}$-test. A next step could be the expansion of the fit model with an intensity scaling that varies with the focal slice in order to take into account effects of photobleaching and illumination intensity fluctuations. Another inroad is to include apodization or amplitude aberrations, which model the dependence of the transmission of the different optical components (objective lens, dichroics) on pupil position. The optimum number of fit parameters can be assessed with statistical methods for model selection, in particular methods based on optimizing the Akaike information criterion, basically a weighted sum of the number of fit parameters and the maximum log-likelihood obtained by the fitting routine \cite{Akaike1974}.

Another next step in the technical developments described in this paper would be to fit only the 3D-position keeping the emission wavelength fixed to the known weighted average fluorescence emission wavelength. This could possibly further improve the $z$-accuracy and precision. The non-zero spectral width could be taken into account by averaging the single-wavelength PSF over the spectrum weighted by the fluorescence emission and dichroic bandpass filter. Such an approach would enable identification of fluorescent labels with different emission spectra using a Generalized Likelihood Ratio Test (GRLT) \cite{Smith2015} by comparing the likelihood of the fit for the known peak wavelengths of the different fluorescent labels.

An improvement of the current experimental setup for photon starved applications would be the replacement of the polarization sensitive LCOS-SLM by a polarization insensitive adaptive optical element such as a Deformable Mirror (DM). A similar calibration and alignment protocol as described here could be used, including the use of a calibration branch to suppress effects of thermal drift in the DM. Another direction of future research is to use the SLM or DM to mitigate the effects of sample induced aberrations. This can be done most effectively when the aberrations primarily originate from a sufficiently thin layer in the sample volume (on the order of the focal depth) by placing the adaptive optical element at a plane conjugate to the aberration inducing layer in the sample \cite{Mertz2015}.

Data and software for aberration retrieval and `3D+$\lambda$' PSF fitting is available at \cite{Stallinga2018code}.

\section*{Funding} 
European Research Council (grant no. 648580); Merton College, Oxford, United Kingdom (Junior Research Fellowship C.S.S.).

\section*{Acknowledgments} 
We thank Aur\`{e}le Adam and Bernd Rieger for useful research advice.




\begin{thebibliography}{99}
\bibitem{Betzig2006}	E. Betzig, G. H. Patterson, R. Sougrat, O. W. Lindwasser, S. Olenych, J. S. Bonifacino, M. W. Davidson, J. Lippincott-Schwartz, and H. F. Hess, ``Imaging intracellular fluorescent proteins at nanometer resolution,'' Science {\bf 313}, 1643--1645 (2006).
\bibitem{Rust2006}	M. J. Rust, M. Bates, and X. Zhuang, ``Sub-diffraction-limit imaging by stochastic optical reconstruction microscopy (STORM),'' Nat. Methods {\bf 3}, 793--795 (2006).
\bibitem{Hess2006} S. T. Hess, T. P. Girirajan, and M. D. Mason, ``Ultra-high resolution imaging by fluorescence photoactivation localization microscopy,'' Biophys. J. {\bf 91}, 4258--4272 (2006).
\bibitem{Hell2007} S. W. Hell, ``Far-field optical nanoscopy,'' Science {\bf 316}, 1153-–1158 (2007). 
\bibitem{Nieuwenhuizen2013} R. P. J. Nieuwenhuizen, K. A. Lidke, M. Bates, D. Leyton Puig, D. Gr\"{u}nwald, S. Stallinga, and B. Rieger, ``Measuring image resolution in optical nanoscopy,''
Nat. Methods {\bf 10}, 557--562 (2013).
\bibitem{Kao1994} H. P. Kao, and A. S. Verkman, ``Tracking of single fluorescent particles in three dimensions: use of cylindrical optics to encode particle position,'' Biophys. J. {\bf 67}, 1291--1300 (1994).
\bibitem{Holtzer2007}	L. Holtzer, T. Meckel, and T. Schmidt, ``Nanometric three-dimensional tracking of individual quantum dots in cells,'' Appl. Phys. Lett. {\bf 90}, 053902 (2007).
\bibitem{Huang2008} B. Huang, W. Wang, M. Bates, and X. Zhuang, ``Three-dimensional super-resolution imaging by stochastic optical reconstruction microscopy,'' Science {\bf 319}, 810--813 (2008).
\bibitem{Toprak2007} E. Toprak, H. Balci, B. Blehm, H. Benjamin, and P.R. Selvin, ``Three-dimensional particle tracking via bifocal imaging,'' Nano Lett. {\bf 7}, 2043--2045 (2007).
\bibitem{Ram2008} S. Ram, P. Prabhat, J. Chao, E. S. Ward, and R. J. Ober, ``High accuracy 3D quantum dot tracking with multifocal plane microscopy for the study of fast intracellular dynamics in live cells,'' Biophys. J. {\bf 95}, 6025--6043 (2008).
\bibitem{Juette2008} M. F. Juette, T. J. D Gould, M. D. Lessard, M. J. Mlodzianoski, B. S. Nagpure, B. T. Bennett, S. T. Hess, and J. Bewersdorf, ``Three-dimensional sub--100 nm resolution fluorescence microscopy of thick samples,'' Nat. Methods {\bf 5}, 527--529 (2008).
\bibitem{Shechtman2014} Y. Shechtman, S. J. Sahl, A. S. Backer, and W. E. Moerner, ``Optimal point spread function design for 3D imaging,'' Phys. Rev. Lett. {\bf 113}, 133902 (2014).
\bibitem{Lew2011} M. D. Lew, S. F. Lee, M. Badieirostami, and W. E. Moerner, ``Corkscrew point spread function for far-field three-dimensional nanoscale localization of pointlike objects,'' Opt. Lett. {\bf 36}, 202--204 (2011).
\bibitem{Pavani2008} S. R. P. Pavani and R. Piestun, ``High-efficiency rotating point spread functions,'' Opt. Express {\bf 16}, 3484--3489 (2009).
\bibitem{Pavani2009} S. R. P. Pavani, M. A. Thompson, J. S. Biteen, S. J. Lord, N. Liu, R. J. Twieg, R. Piestun, and W. E. Moerner, ``Three-dimensional, single-molecule fluorescence imaging beyond the diffraction limit by using a double-helix point spread function,'' Proc. Natl. Acad. Sci. U.S.A. {\bf 116}, 2995--2999 (2009).
\bibitem{Grover2010} G. Grover, S. R. P. Pavani, and R. Piestun, ``Performance limits on three-dimensional particle localization in photon-limited microscopy,'' Opt. Lett. {\bf 35}, 3306--3308 (2010).
\bibitem{Prasad2013} S. Prasad, ``Rotating point spread function via pupil-phase engineering,'' Opt. Lett. {\bf 38}, 585--587 (2013).
\bibitem{Roider2014} C. Roider, A. Jesacher, S. Bernet, and M. Ritsch-Marte, ``Axial super-localisation using rotating point spread functions shaped by polarisation-dependent phase modulation,'' Opt. Express {\bf 22}, 4029--4037 (2014).
\bibitem{Shechtman2015} Y. Shechtman, L. E. Weiss, A. S. Backer, S. J. Sahl,  and W. E. Moerner, ``Precise 3D scan-free multiple-particle tracking over large axial ranges with tetrapod point spread functions,'' Nano Lett. {\bf 15}, 4194--4199 (2015).
\bibitem{Baddeley2011} D. Baddeley, M. B. Cannell, and C. Soeller, ``Three-dimensional sub-100 nm super-resolution imaging of biological samples using a phase ramp in the objective pupil,'' Nano Research {\bf 4}, 589--598 (2011).
\bibitem{Jia2014} S. Jia, J. C. Vaughan, and X. Zhuang, ``Isotropic three-dimensional super-resolution imaging with a self-bending point spread function,'' Nat. Photonics {\bf 8}, 302--306, (2014).
\bibitem{Shtengel2009} G. Shtengel, J. A. Galbraith, C. G. Galbraith, J. Lippincott-Schwartz, J. M. Gillette, S. Manley, R. Sougrat, C. M. Waterman, P. Kanchanawong, M. W. Davidson, R. D. Fetter, and H.F. Hess, ``Interferometric fluorescent super-resolution microscopy resolves 3D cellular ultrastructure,'' Proc. Natl. Acad. Sci. U.S.A. {\bf 106}, 3125--3130 (2009).
\bibitem{Backlund2014} M. P. Backlund, M. D. Lew, A. S. Backer, S. J. Sahl, and W. E. Moerner, ``The role of molecular dipole orientation in single-molecule fluorescence microscopy and implications for super-resolution imaging,'' Chem. Phys. Chem. {\bf 15}, 587--599 (2014).
\bibitem{Broeken2014} J. Broeken, B. Rieger, and S. Stallinga,  ``Simultaneous measurement of position and color of single fluorescent emitters using diffractive optics,'' Opt. Lett. {\bf 39}, 3352--3355 (2014).
\bibitem{Smith2016} C. Smith, M. Huisman, M. Siemons, D. Gr{\"u}nwald, and S. Stallinga, ``Simultaneous measurement of emission color and 3D position of single molecules,'' Opt. Express {\bf 24}, 4996--5013 (2016).
\bibitem{Shechtman2016} Y. Shechtman, L. E. Weiss, A. S. Backer, M. Y. Lee, and W. E. Moerner, ``Multicolour localization microscopy by point-spread-function engineering,'' Nat. Photonics {\bf 10}, 590--594 (2016).
\bibitem{Stallinga2010} S. Stallinga and B. Rieger, ``Accuracy of the Gaussian point spread function model in 2D localization microscopy,'' Opt. Express {\bf 18}, 24461--24476 (2010).
\bibitem{Gibson1992} S. F. Gibson and F. Lanni, ``Experimental test of an analytical model of aberration in an oil-immersion objective lens used in three-dimensional light microscopy,'' J. Opt. Soc. Am. A {\bf 9}, 154--166 (1992).
\bibitem{Stallinga2012} S. Stallinga and B. Rieger, ``Position and orientation estimation of fixed dipole emitters using an effective Hermite point spread function model,'' Opt. Express {\bf 20}, 5896--5921 (2012).
\bibitem{Kirshner2013} H. Kirshner, C. Vonesch, and M. Unser, ``Can localization microscopy benefit from approximation theory?,'' 10th International Symposium on Biomedical Imaging, 588--591 (2013).
\bibitem{Tahmasbi2015} A. Tahmasbi, E. S. Ward, and R. J. Ober, ``Determination of localization accuracy based on experimentally acquired image sets: applications to single molecule microscopy,'' Opt. Express {\bf 23}, 7630--7652 (2015).
\bibitem{Babcock2017}	H. P. Babcock, and X. Zhuang, ``Analyzing single molecule localization microscopy data using cubic splines,'' Sci. Rep. {\bf 7}, 552 (2017).
\bibitem{Li2017} Y. Li, M. Mund, P. Hoess, U. Matti, B. Nijmeijer, V. J. Sabinina, J. Ellenberg, I. Schoen, and J. Ries, ``Fast, robust and precise 3D localization for arbitrary point spread functions,'' bioRxiv 172643; doi: \url{https://doi.org/10.1101/172643}.
\bibitem{Diezman2015} A. Diezmann, M. Y. Lee, M. D. Lew, and W. E. Moerner, ``Correcting field-dependent aberrations with nanoscale accuracy in three-dimensional single-molecule localization microscopy,'' Optica {\bf 2}, 985--993 (2015).
\bibitem{Debarre2007} D. D{\'e}barre, M. J. Booth, and T. Wilson, ``Image based adaptive optics through optimisation of low spatial frequencies,'' Opt. Express, {\bf 15}, 8176--8190 (2007).
\bibitem{Booth2015} M. J. Booth, D. Andrade, D. Burke, B. Patton, and M. Zurauskas, ``Aberrations and adaptive optics in super-resolution microscopy,'' Microscopy {\bf 64}, 251--261 (2015).
\bibitem{Burke2015} D. Burke, B. Patton, F. Huang, J. Bewersdorf, and M. J. Booth, ``Adaptive optics correction of specimen-induced aberrations in single-molecule switching microscopy,'' Optica {\bf 2}, 177--185 (2015).
\bibitem{Hanser2003} B. M. Hanser, M. G. L. Gustafsson, D. A. Agard, and J. W. Sedat, ``Phase retrieval for high-numerical-aperture optical systems,'' Opt. Lett. {\bf 28}, 801--803 (2003).
\bibitem{Liu2013} S. Liu, E. B. Kromann, W. D. Krueger, J. Bewersdorf, and K. A. Lidke, ``Three dimensional single molecule localization using a phase retrieved pupil function,'' Opt. Express {\bf 21}, 29462--29487 (2013).
\bibitem{Kromann2012} E. B. Kromann, T. J. Gould, M. F. Juette, J. E. Wilhjelm, and J. Bewersdorf, ``Quantitative pupil analysis in stimulated emission depletion microscopy using phase retrieval,'' Opt. Lett. {\bf 37}, 1805--1807 (2012).
\bibitem{Petrov2017} P. N. Petrov, Y. Shechtman, and W. E. Moerner, ``Measurement-based estimation of global pupil functions in 3D localization microscopy,'' Opt. Express {\bf 25}, 7945--7959 (2017).
\bibitem{Mortensen2010} K. I. Mortensen, L. S. Churchman, J. A. Spudich, and H. Flyvbjerg, ``Optimized localization analysis for single-molecule tracking and super-resolution microscopy,'' Nat. Methods, {\bf 7}, 377--381 (2010).
\bibitem{Stallinga2015} S. Stallinga, ``Effect of rotational diffusion in an orientational potential well on the point spread function of electric dipole emitters,'' J. Opt. Soc. Am. A {\bf 32}, 213--223 (2015).
\bibitem{Ronzitti2012} E. Ronzitti, M. Guillon, V. de Sars, and V. Emiliani, ``LCoS nematic SLM characterization and modeling for diffraction efficiency optimization, zero and ghost orders suppression,'' Opt. Express {\bf 20}, 17843--17855 (2012).
\bibitem{Persson2012} M. Persson, D. Engstr\"{o}m, and M. Goks\"{o}r, ``Reducing the effect of pixel crosstalk in phase only spatial light modulators,'' Opt. Express {\bf 20}, 22334--22343 (2012).
\bibitem{Lingel2013} C. Lingel, T. Haist, and W. Osten, ``Optimizing the diffraction efficiency of SLM-based holography with respect to the fringing field effect,'' Appl. Opt. {\bf 52}, 6877--6883 (2013).
\bibitem{Akaike1974} H. Akaike, ``A new look at the statistical model identification,'' IEEE Trans. Automatic Control {\bf 19}, 716--723 (1974).
\bibitem{Smith2015} C. S. Smith, S. Stallinga, K. A. Lidke, B. Rieger, and D. Gr\"{u}nwald, ``Probability-based particle detection that enables threshold-free and robust in vivo single molecule tracking,'' Mol. Biol. Cell, mbc.E15-06-044815 (2015). 
\bibitem{Mertz2015} J. Mertz, H. Paudel, and T. G. Bifano, ``Field of view advantage of conjugate adaptive optics in microscopy applications,'' Appl. Opt. {\bf 54}, 3498--3506 (2015).
\bibitem{Stallinga2018code} S. Stallinga, \url{ftp://qiftp.tudelft.nl/stallinga/wavefrontcontrolPSFengineeringSMLM.zip}.
\end{thebibliography}
\end{document}